\documentclass[twocolumn,10pt,a4paper]{article}

\usepackage{graphicx}

\begin{document}

\title{Strength Distribution in Derivative Networks}

\author{Luciano da Fontoura Costa and Gonzalo Travieso\\ 
\small Instituto de F\'{\i}sica de S\~{a}o Carlos, 
       Universidade de S\~{a}o Paulo\\
\small Caixa Postal 369, 13560-970, S\~{a}o Carlos, SP, Brazil\\
\small Phone +55 162 3373 9858, FAX +55 162 3371 3616\\
\small \texttt{\{luciano,gonzalo\}@ifsc.usp.br}}

\maketitle

\begin{abstract}   

This article describes a complex network model whose weights are
proportional to the difference between uniformly distributed
``fitness'' values assigned to the nodes.  It is shown both
analytically and experimentally that the strength density (i.e. the
weighted node degree) for this model, called derivative complex
networks, follows a power law with exponent $\gamma<1$ if the fitness
has an upper limit and $\gamma>1$ if the fitness has no upper limit
but a positive lower limit.  Possible implications for neuronal
networks topology and dynamics are also discussed.

\end{abstract}

\section{Introduction}

Great part of the interest focused on complex networks
\cite{AlbertBarab:2002,Newman:2003,DorogMendes:2002} recently stems
from \emph{scale free} or \emph{power law} distributions of respective
topological measurements, such as the node degree.  At the same time,
weighted complex networks have attracted growing interest because of
their relevance as models of several natural phenomena, with special
attention given to systems used for distribution/collection of
materials or information.

There are two main ways to approach the degrees of weighted networks:
(i) by thresholding the weights and using the traditional node degree
\cite{AlbertBarab:2002}; and (ii) by adding the weights of the edges
attached to each node, yielding the respective node \emph{strength}
\cite{Barthelemyetal:2004}.  Related recent works include the
identification of strength power law in word association networks
\cite{Costawhat:2003,Costahier:2004}, studies of scientific
collaborations and air-transportation networks
\cite{Barratetal:2003}, the analysis of amino acid sequences in terms
of weighted networks \cite{Costahier:2004}, the investigation of
weighted networks defined by dynamical coupling between topology and
weights \cite{Barthelemyetal:2004}, analytical characterization of
thresholded networks \cite{Masudaetal:2004}, as well as the
characterization of motifs \cite{Onnelaetal:2004} and shortest paths
in weighted networks \cite{NohRieger:2002}.

The current article describes a network whose nodes have a respective
``fitness'' value and every node is connected to all other nodes with
higher fitness through an edge whose weight is determined by the
difference between the fitness of the nodes.  Because of such a
construction, these networks are henceforth called \emph{derivative
complex networks}.  Previous works considering the difference of
fitness values associated to the network nodes include the image
analysis approach reported in~\cite{vision:2004}, which involves the
thresholding of the fitness difference values in order to perform
image segmentation and the gradient networks discussed
in~\cite{Toro_Bassler:2004,Toro_etal:2004}, which take into account
the edges corresponding to the highest differences, i.e. to the
gradient of a scalar field associated along the network nodes.  

The developments in the current article represent a continuation and
extension of preliminary investigations reported
in~\cite{Costa_strength:2004}. We show both analytically and
numerically that the strength densities for the derivative network
model follow a power law with exponent $\gamma<1$ if the fitness has
an upper limit and $\gamma>1$ if the fitness has no upper limit but a
positive lower limit. 

We start by describing the derivative network model and follow by
calculating its respective strength distribution and discussing
possible implications for neuronal networks.

\section{Derivative Networks}

Consider a network with $N$ nodes. To each node $i$, a ``fitness''
value $\varphi_i > 0$ is randomly assigned following a distribution
$\rho(\varphi)$. The fitness derivative determines the connectivity of
the nodes in the network. A directed arc linking node $j$ to node $i$
is drawn iff $\varphi_i > \varphi_j$; in that case, the arc has weight
given by a function of the fitness difference:
\begin{equation}
  w_{ji} =\sigma(\varphi_i-\varphi_j).
  \label{weight}
\end{equation}
We are here interested in the study of the distribution of input
strengths in the network under some assumptions on the functions
$\rho$ and $\sigma$. The input strength of node $i$, $s_i$, is the sum
of the weights of the arcs linked to $i$:
\begin{equation}
  s_i = \sum_{j} w_{ji},
  \label{strength}
\end{equation}
where the sum is taken over all nodes $j$ that have an arc linked to
node $i$.

From Eqs.~(\ref{strength}) and~(\ref{weight}), the strength of a node
is determined by its fitness $\varphi$:
\begin{equation}
  s(\varphi) = \sum_{\varphi' < \varphi} \sigma(\varphi-\varphi');
\end{equation}
where the sum is taken for all nodes with $\varphi'$ smaller than
$\varphi$.  Considering the distribution of fitness $\rho(\varphi)$,
for the limit of large number of nodes $N$ we can write:
\begin{equation}
  s = N \int_{0}^{\varphi} \rho(x) \sigma(\varphi-x)\,dx.
  \label{sxq-general}
\end{equation}

For the weight function, we assume a power law as:
\begin{equation}
  \sigma(x) = b x^{\beta-1}.
  \label{sigma}
\end{equation}
We are interested in the case where the weight grows with the
difference in fitness, implying $\beta > 1.$

For the fitness distribution, a power law distribution will be also
considered. In the following sections, distributions with an upper
limit and with a positive lower limit will be considered.

\subsection{Upper bound fitness}

First we study the case where the values of fitness have a maximum
value that we assume, without loss of generality, to be unitary: $0 <
\varphi < 1.$ For a power law distribution
\begin{equation}
  \rho(\varphi) = \alpha \varphi^{\alpha-1}
  \label{rhoupper}
\end{equation}
to be a valid probability distribution in the interval $0 < \varphi <
1,$ we must have $\alpha > 0.$ From Eqs.~(\ref{sxq-general}),
(\ref{sigma}) and~(\ref{rhoupper}) we have:
\begin{equation}
  s = N \int_{0}^{\varphi} \alpha x^{\alpha-1} b
(\varphi-x)^{\beta-1}\,dx, 
\end{equation}
resulting in:
\begin{equation}
  s = N b \alpha \mathrm{B}(\alpha,\beta) \varphi^{\alpha+\beta-1},
  \label{sxq-upper}
\end{equation}
where $\mathrm{B}(a,b)$ is the Euler beta function:
$$ \mathrm{B}(a,b) = \int_{0}^{1}t^{a-1}(1-t)^{b-1}\,dt. $$
Note that this expression implies an upper limit for $s$ given by
\begin{equation}
  s_\mathrm{max} = N b \alpha \mathrm{B}(\alpha,\beta).
  \label{smax}
\end{equation}

The probability density of $s$, $p(s)$, is given by:
\begin{equation}
  p(s) = \rho(\varphi)\frac{d\varphi}{ds}.
  \label{pspphi}
\end{equation}
The derivative can be obtained from Eq.~(\ref{sxq-upper}), giving:
\begin{equation}
  p(s) = \frac{\alpha}{\alpha+\beta-1}
   \left(\frac{1}{s_\mathrm{max}}\right) ^
    {\frac{\alpha}{\alpha+\beta-1}}
    s^{-\frac{\beta-1}{\alpha+\beta-1}}.
  \label{ps-upper}
\end{equation}
This expression is only valid for $0 \le s < s_\mathrm{max},$ and is a
power law $p(s) \sim s^{-\gamma_U}$ with 
\begin{equation}
  \gamma_U = \frac{\beta-1}{\alpha+\beta-1}.
  \label{gammau}
\end{equation}

Considering $\alpha>0$ and $\beta>1$ we have $0 < \gamma < 1.$

An interesting special case is that of uniform distribution, which
corresponds to $\alpha=1$ in Eq.~(\ref{rhoupper}), resulting in
$$ \gamma_U = 1-\frac{1}{\beta}.$$ In case we also have that the
weight is proportional to the fitness difference, i.e.\ $\beta=2$ in
Eq.~(\ref{sigma}), then $\gamma_U = \frac{1}{2}.$

\subsection{Lower bound fitness}

We consider now the case where the fitness value has no upper limit,
but a positive lower limit, that we assume, without loss of
generality, to be unitary. In this case, we write the probability
density as
\begin{equation}
  \rho(\varphi) = \alpha \varphi^{-\alpha-1},
  \label{rholower}
\end{equation}
which is correctly normalized in the interval $1 < \varphi < \infty$
for $\alpha>0.$

Using Eqs.~(\ref{sxq-general}), (\ref{sigma}), and~(\ref{rholower}) 
we have:
\begin{equation}
  s = N \int_{1}^{\varphi} \alpha x^{-\alpha-1} b
(\varphi-x)^{\beta-1}\,dx, 
\end{equation}
giving:
\begin{equation}
  s = N b \alpha \mathrm{B}(\frac{1}{\varphi},1,-\alpha,\beta)
      \varphi^{\beta-1-\alpha}.
  \label{sxq-lower}
\end{equation}
where $\mathrm{B}(x,y,a,b)$ is the generalized incomplete beta function:
$$ \mathrm{B}(x,y,a,b) = \int_{x}^{y}t^{a-1}(1-t)^{b-1}\,dt. $$

For $\varphi\approx 1,$ $\mathrm{B}(\frac{1}{\varphi},1,-\alpha,\beta)
\approx 0$ giving $s \approx 0.$ For large
$\varphi,$ $\mathrm{B}(\frac{1}{\varphi},1,-\alpha,\beta)
\sim  \varphi^{\alpha}.$ Then
\begin{equation}
  N b \alpha \mathrm{B}(\frac{1}{\varphi},1,-\alpha,\beta) \simeq
  c \varphi^\alpha
  \label{approxbeta}
\end{equation}
and therefore we get:
\begin{equation}
  s \simeq c \varphi^{\beta-1}
  \label{sxq-lower2}
\end{equation}
for large $\varphi.$

Proceeding in a similar fashion to the previous case we get:
\begin{equation}
  p(s) \simeq \frac{\alpha}{\beta-1}
    {c^{\frac{\alpha}{\beta-1}}}
    s^{-\frac{\beta-1+\alpha}{\beta-1}},
  \label{ps-lower}
\end{equation}
valid for large $s.$

Considering that $\beta>1$ and $\alpha>0,$ this expression corresponds
to a power law $p(s) \sim s^{-\gamma_L}$ with $\gamma_L > 1.$

\section{Simulation results}

\begin{figure}
  \begin{center}
  \includegraphics{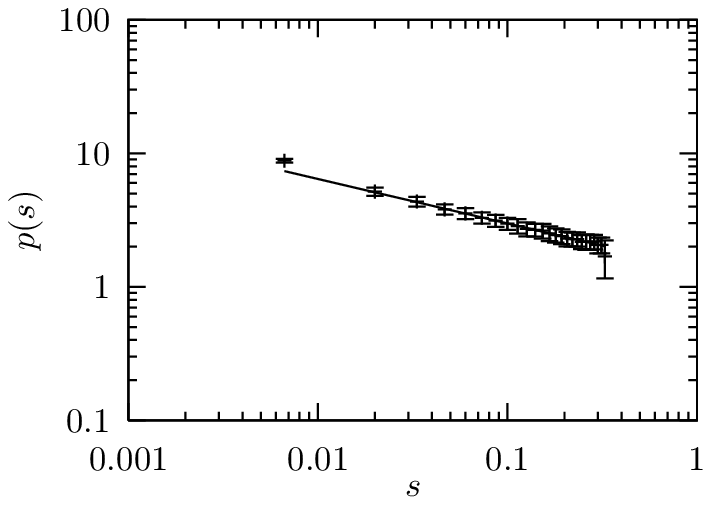}\\
  (a)\\
  \includegraphics{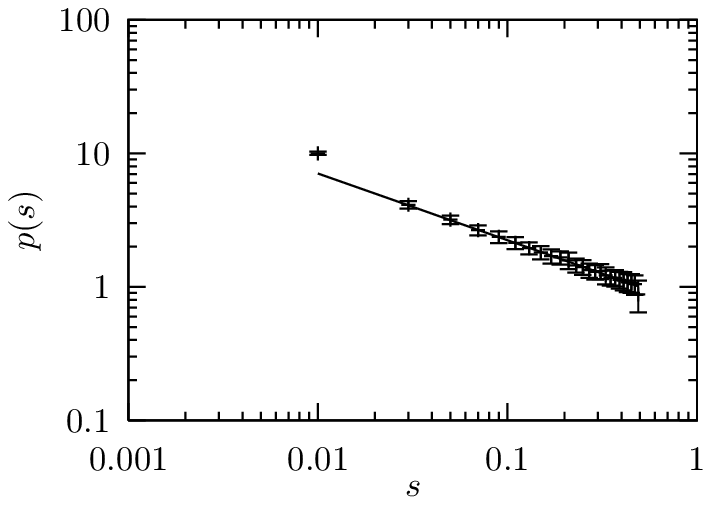}\\
  (b)\\
  \includegraphics{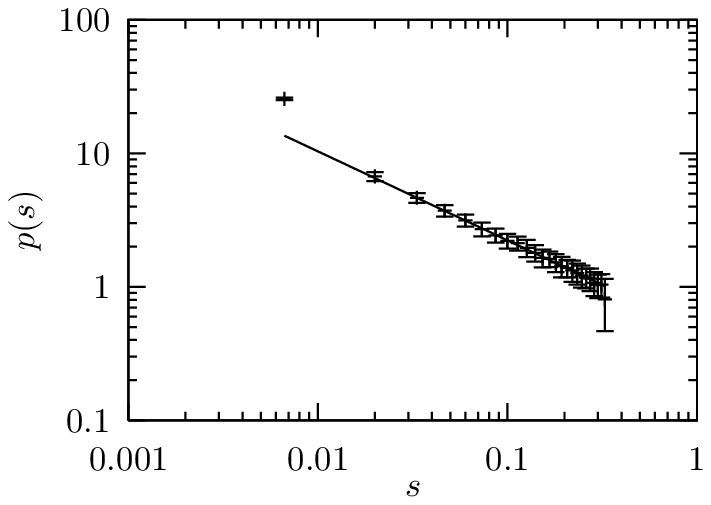}\\
  (c)
  \end{center}

  \caption{Strength distribution in upper limited networks.
    Experimental results are represented by points with error bars
    (one standard deviation) and the continuous curves are the
    theoretical expressions. Results for:
    (a)~$\alpha=2,\beta=2$ ($\gamma=1/3$); 
    (b)~$\alpha=1,\beta=2$ ($\gamma=1/2$); (c)~$\alpha=1,\beta=3$
    ($\gamma = 2/3$).}
  \label{fig:psu}
\end{figure}

\begin{figure}
  \begin{center}
  \includegraphics{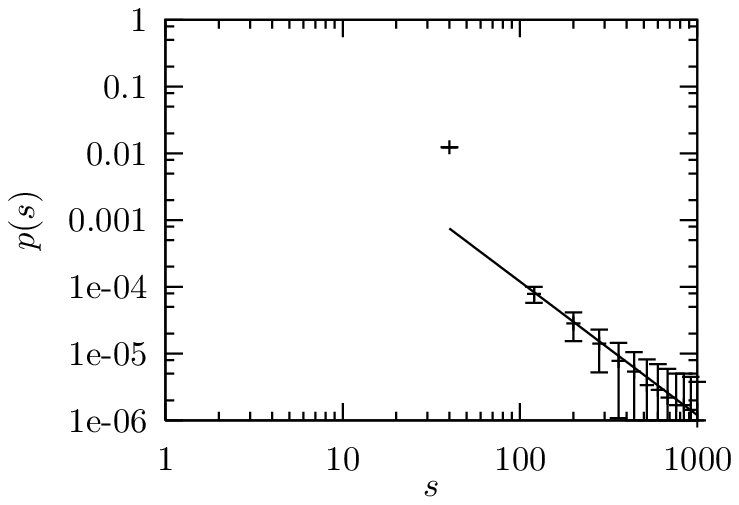}\\
  (a)\\
  \includegraphics{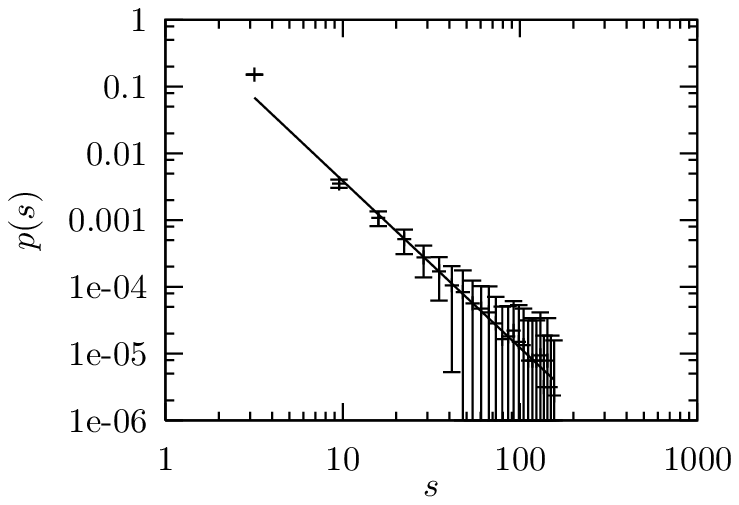}\\
  (b)\\
  \includegraphics{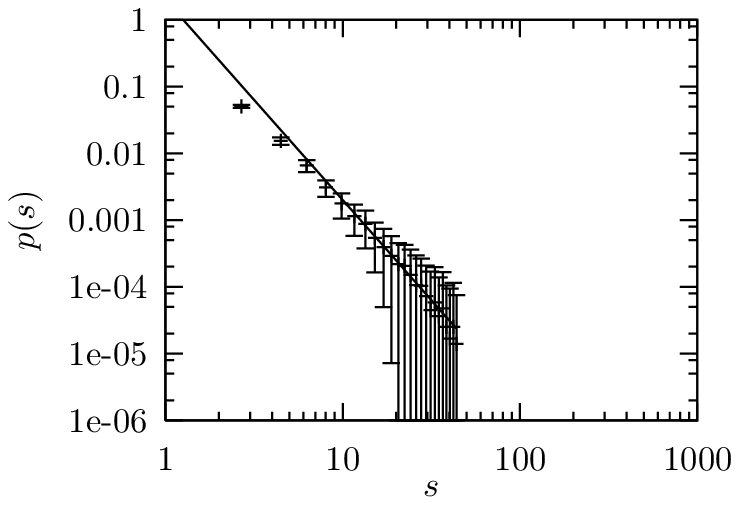}\\
  (c)
  \end{center}

  \caption{Strength distribution in lower limited networks. 
    Experimental results (points with error bars) are represented
    together with a continuous line displaying the theoretical
    asymptotic inclination. Results for:
    (a)~$\alpha=0.5,\beta=1.5$ ($\gamma=2$); (b)~$\alpha=1.5,\beta=2$
    ($\gamma=2.5$); (c)~$\alpha=1,\beta=1.5$ ($\gamma=3$). }  
  \label{fig:psl}
\end{figure}

We simulated derivative networks with upper and lower limit for three
sets of $\alpha$ and $\beta$ parameters each. Figure~\ref{fig:psu}
shows the strength distribution for networks with upper fitness limit;
figure~\ref{fig:psl} shows the results for networks with lower limited
fitness. The results assumed $N=2000$ nodes and $b = 0.0005$; a
total of $100$ networks were considered for each set of parameters, and
the results show the mean values and respective standard deviations.

The curves obtained by the simulations resulted remarkably close to
the analytical results for upper limited networks, except for the
smallest strengths. This is because the strength distribution $p(s)
\sim s^{-\gamma}$ approaches a discontinuity at zero.  

Less adherence between theoretical and experimental values was
observed for the lower limited networks as a consequence of the fact
that the approximation in Eq.~(\ref{approxbeta}) is only valid for large
values of $s,$ which corresponds to small values of the probability
$p(s),$ resulting in poor statistics.

\section{Discussion}

We have described two simple models of derivative networks and shown
that their node instrength distributions follow a power law under some
assumptions on distribution of fitness and weight assignment.
This result implies that although several nodes of a derivative
network have low strength values, there are hubs characterized by a
variety of large instrength values.  Because the instrength of a node
can be understood as the total weighted influence it receives from the
adjacent nodes, it follows that in case the node state is directly
proportional to its instrength, the linear dynamics of such networks
will be characterized by a near power law with similar parameters as
those of the instrength density.  Such an interpretation is
particularly interesting from the perspective of integrating neuronal
network and complex network research (e.g.~\cite{Stauffer:2003}),
where each neuron is represented by a node and the synaptic
connections by directed edges whose weights reflect the respective
synaptic strengths.  The ``fitness'' values in this case could be
related to gradient of neurotrophic growth factors or depolarization
bias facilitating the action potential
\cite{Kandel:1995,Freeman:2001}.  Another approach worth further
attention would be to consider derivative networks as models of
perceptual and cognitive processes, which are known to explore
derivatives as a means of diminishing stimuli redundancy
(e.g.~\cite{Schiffman}).

\section*{Acknowledgments}
Luciano da F. Costa is grateful to FAPESP (process 99/12765-2), CNPq
(308231/03-1) and the Human Frontier Science Program (RGP39/2002) for
financial support.

\bibliographystyle{unsrt}
\bibliography{powerr-ex}

\end{document}